# Exploration of Strange Electromagnetics in Carbon Films

S. G. Lebedev

Institute for Nuclear Research of Russian Academy of Sciences,
60th October Anniversary Prospect,7a, Moscow
117312, Russia, tel.:70953340189, fax:70951352268,
e-mail: lebedev@inr.ru

Results of magnetic force microscopy (MFM), dc SQUID magnetization, reversed Josephson effect (RJE), and resistance measurements in thin carbon arc (CA) films are presented. The observation of a RJE induced voltage as well as its rf frequency, input amplitude, and temperature dependence reveals the existence of Josephson Junction arrays. Oscillating behavior of the DC SQUID magnetization reminiscent of the Fraunhofer-like behavior of superconducting (SC) critical current in the range of 10000 Oe has been observed. The dc SQUID magnetization measurement indicates a possible elementary 102 nm SC loop; this is compared to MFM direct observations of magnetic clusters with a median size of 165 nm. All these data are consistent with the existence of a high temperature SC-like phase or fluctuations up to 650 K. It is proposed to expose such CA film to energetic particle (neutron or ion) bombardment to verify this hypothesis. Such bombardment would change both the structure of film and consequently the experimental measurements. In addition such bombardment-induced changes will provide a basis for particle detectors utilizing the Josephson effect.

PACS numbers: 74.25.H; 74.76; 85.25.C. Key words: thin films; superconductivity; Josephson devices.

## I. INTRODUCTION

Recently superconducting-like behavior was found in highly oriented pyrolitic graphite (HOPG) [1-3]. Also some weak superconducting behavior has been observed in carbon films formed by sputtering of spectroscopically pure graphite in an electrical arc discharge (CA films) [4-5]. In both of these two materials there is some paramagnetic response in high field magnetization measurements. Such behavior in graphite cannot be adequately explained by ferromagnetic impurities [1]. In the case of the CA films, the absence of unpaired spins and paramagnetic signal has been shown in earlier ESR experiments[4]. On the other hand the existence of superconducting-like hysteresis loops in HOPG [1], reversed Josephson effect (RJE) in CA films [4] still suggest the coexistence of superconductivity (SC) and paramagnetism (PM) in pointed materials. This behavior is similar to high-temperature SC when they are cooled through their transition temperature in a small external magnetic field. This result, the paramagnetic Meissner effect (PME), contrasts with the standard diamagnetic response of classical superconductors and has been the subject of extensive investigations for the last years. An experimental study of the paramagnetic response in Bi-based polycrystalline SC was first made by Braunisch et al. [6] and subsequently reported in many other high-temperature superconductor (HTSC) [7-10].

Some theoretical work has suggested that PME might be attributed to the magnetic impurity between the grains forming the anomalous Josephson junctions ($\pi$-junctions)[11] or non-s-wave pairing symmetry [12]. Still other work suggests the topological disorder enhances the density of state in graphene sheets and can give rise to coexistence of SC and PM fluctuations [13].

A common feature of all these systems is their granular structure forming the Josephson-junction arrays (JJA). Therefore a lot of modeling experiments have been conducted to check the existence of PME in JJA. Dominguez et al.[14] have modeled a granular SC as a mixture of conventional and $\pi$-junctions network where the nodes represent the grains and the links represent the coupling between them. In this case, the low-temperature and zero-field-cooling (ZFC) magnetic susceptibility was of the order of

-1 (in SI units), while the field-cooling (FC) susceptibility was paramagnetic for some small value of magnetic field. Other models based on networks of conventional junctions could explain the PME [15-18]. In one case [17] the JJA of conventional junctions has been used with an elementary loop of size of 46 μm consisting of four Josephson junctions. In the ac susceptibility measurements up to the magnetic field $h_{ac}$ of about 50 mOe **(0.4 amp/m)**, which corresponds to $5\Phi_0$ per elementary loop, the diamagnetic signal has been observed. Upon increasing $h_{ac}$ above 50 mOe **(0.4 amp/m)** the PME reappears. As it has been shown this non-monotonic behavior of magnetisation is the consequence of a magnetic field dependence of critical current in elementary loop (Fraunhofer pattern).

The CA films and HOPG materials differ from high-temperature SC and their JJA model in the scale of magnetic field. In our case the magnetic field where the PME occurs is about 1T as compared with the 50 mOe in the experiments of Refs. [17-18]. However the MFM study reveals the size of elementary conductive loop of granules in CA films being about 500 times smaller than in HTSC yet having the same magnetic flux per SC loop. So it is believed we can proceed our measurements and analysis partly in a fashion similar to Refs.[15-18].

This paper presents the results of dc magnetisation, magnetic force microscope, low and high temperature electrical resistance measurements, and RJE voltage- all on CA films. Each of these results cannot be explained unambiguously, but considered together they suggest the existence of SC phase or fluctuation in CA films. Of course these experiments can be considered only as preliminary and should be checked carefully. However independent of the explanation of these findings, CA films can be used in applications as Josephson detectors.

## II. EXPERIMENTS

The CA films for our measurements were prepared by arc evaporation of 99.999% purity carbon onto quartz substrates at room temperature [4]. Potassium-oleate ($C_{18}H_{33}O_2K$) was used as the release agent. Self-supporting films were floated from the substrate using distilled water. The CA films were annealed at 1000 C for 10 hours [4].

The thickness of annealed films was about 960 nm. 2 MeV $H^+$-PIXE analysis indicated that the Fe concentration in the annealed films was 185 ± 38 ppm, and the film density was about 2.25 g/cm$^2$.

The temperature dependence of resistance was measured in a commercial closed cycle refrigerator and in a vacuum furnace, below (20 K< T <325 K) and above (300 K < T < 777 K) room temperature, respectively. In the low temperature resistance measurements, a Lakeshore 330 temperature controller digitally recorded the sample temperature using a RS-Components PT-100 platinum resistor placed directly on the sample. With this cryogenic system, samples could be cooled down to 20 K in about 2 h. Warming up of the system took longer, up to 10 h to go from the lowest temperature to room temperature. A Keithley 220 programmable current source was used to supply a current , I, which was also monitored by a four-wire precision resistor of 0.1Ω. The voltage drop, V, on this resistor and on the sample was measured with a Keithley 2182 Nanovoltmeter. An Agilent 34970A Switch Unit, with a 34901A 20 Channel Multiplexer card and a 34970A 4×8 Matrix Switch card were used to allow for switching of both voltages and the current polarity, respectively, interconnecting the test leads with the nanovoltmeter and the current source in a break-before-make mode. All these instruments had IEEE-488.2 interfaces and used an IBM-compatible personal computer as the controller. The control software, written in LabVIEW from National Instruments, programmed a current value and selected the current polarity between the source and the sample. Therefore both voltage drop, V, the sample temperature, T, and the time, t, were digitally recorded. All measurements were continuously stored in ASCII-files while the sample was either cooled down or warmed up.

To provide the four-probe electrical resistance measurements, thin gold wire electrical contacts were attached to the film samples with the help of silver paste. Usually contacts had dimensions of 3 mm by 1 mm.

MFM measurements were employed to study the topography of film surfaces and possible magnetic clusters. Magnetic force gradient images and sample topography were obtained simultaneously with a Nanoscope III scanning probe microscope from Digital Instruments. The microscope was operated in the "tapping/lift ™" scanning mode, to separate short-range topographic effects from any long-range magnetic signal. The scanning probes were batch fabricated Si cantilevers with pyramidal tips coated with a magnetic CoCr film alloy. Prior to acquiring an image, the probe was exposed to a 1-3 kOe permanent magnet, which aligned its magnetization normal to the sample surface direction (i.e., z direction). All MFM data shown in this paper were collected with the tip magnetized nearly perpendicular to the sample surface, making the MFM sensitive to the second derivative of the z component of the sample stray field. Images were taken with various tip-sample separations (10 ~ 100 nm) with the requirement that the general shape of MFM images would not change with the tip-sample distance variation. This excluded any influence of the MFM tip on the sample micromagnetic structure and verified that any non-z components of the tip magnetization contributed negligibly to the MFM measurements,.

## III. RESULTS AND DISCUSSION

Figure 1 shows the temperature dependence of electrical resistance in annealed CA films. The resistance smoothly decreases as temperature increases from 20 K to 777 K. This tendency is reproducible with warming up and cooling down. The change of resistance with temperature from room temperature to 777 K is about 17 ~ 25 %. Some of the CA samples - in contrast with the common behavior of Fig.1 - showed a change of slope or even a growth of resistance in the temperature range T >650-700 K. However these results were generally not reproducible.

In order to check the existence of JJA in our carbon arc films, we performed the measurements of RJE. We have studied the RJE in various experimental arrangements of Refs. [22] and [23] and measured the induced dc voltage, $V_{dc,}$ as a function of frequency, temperature, and input ac amplitude. The fluctuations of $V_{dc}$ are fast and it may be the result of a series combination of a large number of individual Josephson junctions. The dependence of $V_{dc}$ on the rf frequency, f, at room temperature is shown in Fig. 2. In this measurement, we used the same circuit as in Ref. [22]. Note that there is a polarity reversalof $V_{dc}$, which is one of the tests which can distinguish the RJE from the commonly known rectification effect [22]. Moreover, in our CA films, there is a hysteresis in $V_{dc}$ with frequency as shown in figure 2, i.e. $V_{dc}(f)$ differs depending on whether frequency was increasing or decreasing for some high frequency region i.e., $f > 9$ MHz. This hysteresis was not found when we used the circuit of Ref. [23].

Figure 3 shows the change of $V_{dc}$ with the input ac amplitude, $V_{ac}$ for several fixed rf frequencies by using the same circuit as in Ref. [23]. For a small rf frequency region ($f < 1.6$ MHz), $V_{dc}$ increases with input ac amplitude and has a positive sign. However at $f = 1.6$ MHz it changes polarity (see inset of Fig. 3) and for higher frequencies ($f > 1.6$ MHz,) it is negative and continues to decrease with $V_{ac}$. For a small $V_{ac}$ it is difficult to find the change of polarity, however, for higher $V_{ac}$ it is possible to resolve the polarity change of $V_{dc}$ with rf frequency.

For comparison, we studied the behavior of a standard carbon resistor, with the same resistance as the CA film sample. We did not observe the change of polarity in the standard carbon resistor, and the amplitude of $V_{dc}$ was one order of magnitude smaller than that for CA films. This reveals that the polarity change of $V_{dc}$ is a property peculiar to CA films.

The temperature dependence of $V_{dc}$ was measured by using the electrical circuit of Ref. [23] at fixed input amplitude and rf frequency. At all temperatures, the data were obtained after reaching thermal equilibrium. The result is shown in Fig. 4 for 300 K < T < 777 K at $V_{ac}$ = 10 V and rf frequency f= 1 MHz. $V_{dc}$ exponentially decreases with temperature. This can be successfully fitted by the form $V_{dc}=\alpha e^{bT}$ for 300 K < T < 650 K, where 1/b= -104±60 (K) which is in good agreement with the result of Refs. [4,23]. Around 650 K $V_{dc}$ goes to zero and starts to be negative above 650 K. Also, the change of $V_{dc}$ with temperature does not follow the exponential behavior for T>650 K. From 300 K to 777 K, the change of $V_{dc}$ is more than 100 % and thus larger than the 17 - 25 % change shown in Figure 1.

In fresh films there was evidence of corrugating or warping of the surface in the topographic images and an absence of magnetic clusters or domains in MFM images. Here the scan size was 5 μm x 5 μm and the scan height in MFM measurement was 50 nm. The annealed films had a smooth surface at large scale and revealed the topographic clusters with the average size of about 165 nm (see Fig. 5). As can be seen from the MFM image in Figure 5, it is possible to resolve the magnetic domains or particles. After magnetization of CA film with the use of permanent magnet with the magnetic field of 80 kamp/m there was clear evidence of magnetic clusters with the same sizes and locations (see Fig. 6) as the topographic clusters. This suggests that there were magnetic clusters or domains in the annealed CA films and the magnetization of magnetic domain is small, which however can be resolved in DC SQUID magnetization measurements.

DC magnetization M(H,T) measurements were performed with a SQUID magnetometer MPMS7 from Quantum Design. The M(H) curves for different temperatures are shown in Fig.7. As can be seen from Fig.7 there is some paramagnetic response at H<10000 Oe**.** At H>800 kamp/m the signal is negative due to natural diamagnetism of graphite granules embedded in the matrix of both twofold and fourfold-coordinated atoms [19-20]. After subtraction of this linear diamagnetic background $M= -\chi(T)H$, we found the oscillating behavior of magnetisation (see inset of Fig. 7). This is similar to the dependence of the Josephson junction critical current on a magnetic field applied in the plane of the junction (Fraunhofer pattern). This result is in agreement with the ones obtained in Refs.[16,17,21] with only one difference - in the scale of magnetic field. As is well known the minima of critical current in the Fraunhofer pattern correspond to integral magnetic flux quanta. Therefore for the first minimum in correspondence with Refs.[17-18] we have the relation:

$$5\Phi_0=\mu_0 a^2 H , \qquad (1)$$

where $\Phi_0$ is the magnetic flux quantum, $\mu_0$ is the vacuum permeability, a is the size of elementary SC loop of granules which are in the state of common phase coherence. This means that inside this elementary loop of granules the SC current can flow. The first maximum of M(H) in Fig.7 (which correspond to first minimum of critical current in a Fraunhofer pattern) gives the value of a about 102 nm as compared with the 46 μm in the experiments described in Refs.[17-18]. According to equation (1), this suggests a magnetic field scale $2x10^5$ times larger in our case.

The data of Fig.7 also provide us with the value of critical current $I_c$ for elementary Josephson junctions composing the film granular structure. For this purpose we can use the relation from Ref.[17] which connect the magnetisation with $I_c$:

$$M = \frac{LI_c}{\mu_0 a^2} \quad . \qquad (2)$$

Where L is the inductance of the elementary loop. This gives us the value of $I_c$(100K)~0.6 μA for the annealed film. This value gives the upper evaluation of the critical current density of about $10^4$A/cm$^2$. However taking into account the cross section of current flow between adjacent loops must be substantially smaller than the loop square we can obtain the current density close to the pairs breaking value. It is interesting to note that the critical current in non-annealed film is evaluated to be 30 times larger than in an annealed one. This means that the annealing may suppress the Josephson pairing between the granules. In this sense the observed corrugations and warped nature of the unannealed film surface may turn out to be the same topological disorder which enhances the density of states at the Fermi level of graphite sheets as was shown in Ref. [13].

The size of the elementary loop inferred from the magnetisation experiments is nearly the same as one observed in MFM –measurements (see Fig.5). So it is believed this coincidence is not a casual but is, in fact, the evidence of the existence of SC loops in the CA film samples.

Of course the results obtained cannot be considered as unambiguous evidence of a SC phase or correlation in CA films. On the contrary this is only the first indication of some features in electromagnetic behavior of CA films, which require further investigations to check the results using other equipment and methods. One of such check method may be the study of the influence of energetic particle irradiation on the parameters measured experimentally. These could be the location and amplitude of Fraunhofer-like oscillation of magnetisation, MFM pictures, RJE- related and R(T) data. Asa is well known [24] energetic particles irradiation gives rise to drastic modification of solid-state material structure. This is not only due to point defect generation but also due to displacement cascades occurring into the target media under energetic particles bombardment. It was shown that the cascade core is a vacancy-enriched zone while the cascade periphery is, on the contrary, an interstitial-enriched zone. With rising particle energy the size of cascade zone increases. Depending on particle energy, charge and mass the size of cascade zone may extend from a few up to hundreds and thousands atomic sites in the crystalline lattice. So cascades can modify the solid-state structure on both short and long ranges. This suggests that CA film structure can perhaps be changed controllably by varying particle irradiation parameters such as irradiation dose, charge, or mass and energy of incident particle. Of course such irradiation must be combined with measurement of the main electromagnetic parameters discussed above. Particularly it will be very interesting to observe the shift of Fraunhofer-like oscillations position and change of their amplitude due to, for example, neutron irradiation. Some description of such an experiment with conventional HTSC published in Ref. [25] shows that neutron irradiation up to 0.1 DPA ($10^{19}$n/cm$^2$) gave rise to substantial change of critical current value. Irradiation with a proton beam at the Moscow Meson Facility [24] with today's energy about 300 MeV and proton intensity up to 150 μA can even substantially accelerate this experiment.

In conclusion, it should be noted again these experiments must be considered as preliminary and should be checked carefully. However, independent of the explanation of the observed results, CA films can be used in applications as a Josephson detectors. Some interesting application of CA films could be a non-cryogenic Josephson detector of gamma-radiation for registration of a neutrinos and dark matter [26,27].

## III. SUMMARY

- Experimental results presented in this work have been obtained with the help of such techniques as dc SQUID magnetisation, MFM, RJE, and R(T) measurements. These results support the existence of SC phase or fluctuation in CA films at room and possibly higher temperatures.
- From the dc magnetisation measurements the size of an elementary SC loop of 102 nm, and a critical current in such loop of 0.6 μA has been obtained. This value has been found to be in agreement with MFM measurements.
- The observation of dc voltage induced in samples due to RJE and their frequency and temperature dependencies, which agree well with the previous observations, has been considered as independent proof of existence of JJA and therefore the presence of SC phase or fluctuation inside CA films.

- The disappearance or change of polarity of RJE signal above 650 K reveals that SC phase can persists up to 650 K.
- Energetic neutron or even proton irradiation can help to check the results obtained and shed some light on the origin of electromagnetic features observed in CA films.

## IV. ACKNOWLEDGEMENTS

The author would like to thank Prof. P.Esquinazi, Drs. K.-H.Han, F.Mrowka, R.Hohne and Mrs. A.Setzer from the Department of Superconductivity and Magnetism of Institute of Experimental Physics II of University of Leipzig (Germany) for their help in the experimental measurements. I'd like also to thank Anatoly Shamanin for providing us with the fresh CA films.

## REFERENCES

1. P. Esquinazi et al., J.Low. Temp. Phys. 119(2000)691.
2. P. Esquinazi et al., Fizika Tverdogo Tela 41(1999)2135.
3. H. Kempa et al., Solid State Communications 115(2000)539.
4. S. G. Lebedev and S. V. Topalov, Bulletin of Lebedev's Physical Institute N11-12(1994)14.
5. S. G. Lebedev, Preprint INR RAS, N 1033/2000, Moscow, 2000 (in Russian).
6. W. Braunisch et al., Phys. Rev. Lett. 68(1992)1908.
7. B. Schliepe et al., Phys. Rev. B47(1993)8331.
8. S. Riedling et al., Phys. Rev. B49(1994)13283.
9. U. Onbasli et al., Phys. Status Solidi B194(1996)371.
10. G. S. Okram et al., Condens. Matter 9(1997)L525.
11. F.V. Kusmartsev, Phys.Rev. Lett. 69(1992)2268.
12. H. Kanamura and M. S. Li, Phys. Rev. B54(1996)619.
13. J. Gonzalez et al., Phys. Rev. B63(2001)134421.
14. D. Dominguez et al., Phys. Rev. Lett. 72(1994)2773.

15. C. Auletta et al., Phys. Rev. B51(1995)12844.
16. Mahesh Chandran, Phys. Rev. B56(1997)6169.
17. P. Barbara et al., Phys. Rev. B60(1999)7489.
18. A. P. Nielsen et al., Phys. Rev. B62(2000)14380.
19. L. Salamanca-Riba et al., Nucl. Instr. and Meth. B7/8(1985)487.
20. C. Z. Wang et al., Phys. Rev. Lett., 70(1993)611.
21. W. A. C. Passos et al., J. Appl. Phys. 87(2000)5555.
22. J. T. Chen et al., Phys. Rev. Lett. 58(1987)1972.
23. R. Munger and H. J. T. Smith, Phys. Rev. B44(1991)242.
24. E.A.Koptelov, S.G.Lebedev et al., Nucl.Instr. and Meth. A480(2002)137-155.
25. R.F.Konopleva and V.S.Chashchin, Fizika Tverdogo Tela 39(1997)28-34 (in Russian).
26. L.Stodolsky, Phys. Today, 44(1991)25.
27. A.Da Silva et al., Proc. Of Workshop on Low Temperature Detectors for Neutrinos and Dark Matter II, p.417, France, 1988.

Figure captions

Fig.1. Temperature dependence of electrical resistance in annealed CA films.
Fig.2. Frequency dependence of $V_{dc}$ in CA films using the circuit of Ref.[23].
Fig.3. Change of induced $V_{dc}$ with the input ac amplitude, $V_{ac}$ for several fixed rf frequencies for the same circuit as in Ref. [24].
Inset of Fig.3. Change of polarity of $V_{dc}$ ($V_{ac}$) for f = 1.6 MHz.
Fig.4. Temperature dependence of the RJE induced $V_{dc}$.
Fig.5. Topographic (left) and magnetic (right) clusters in annealed CA film.
Fig.6. Sharp increase of intensity of magnetic clusters shown in Fig.6 after magnetization of CA film with the use of permanent magnet with the magnetic field of H =1kOe.
Fig.7. The magnetization M(H) curves for different temperatures. Linear fits to these curves also shown.
Inset of Fig.7. The oscillating behaviour of magnetisation reminding the Fraunhofer pattern for Josephson junction critical current. This picture has been obtained after subtraction the linear diamagnetic background $M= -\chi(T)H$.

**Figure 1.**

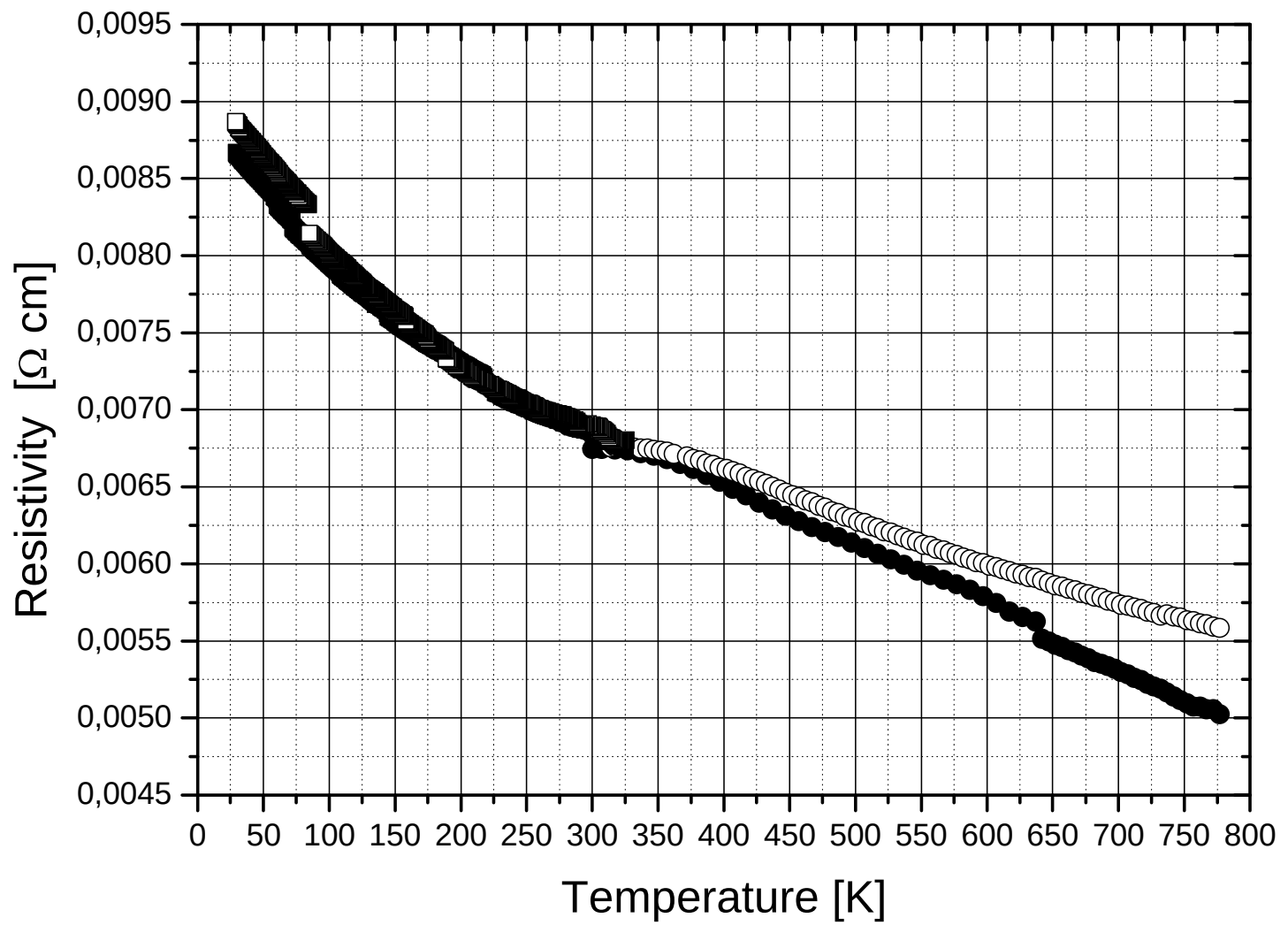

**Figure 2**

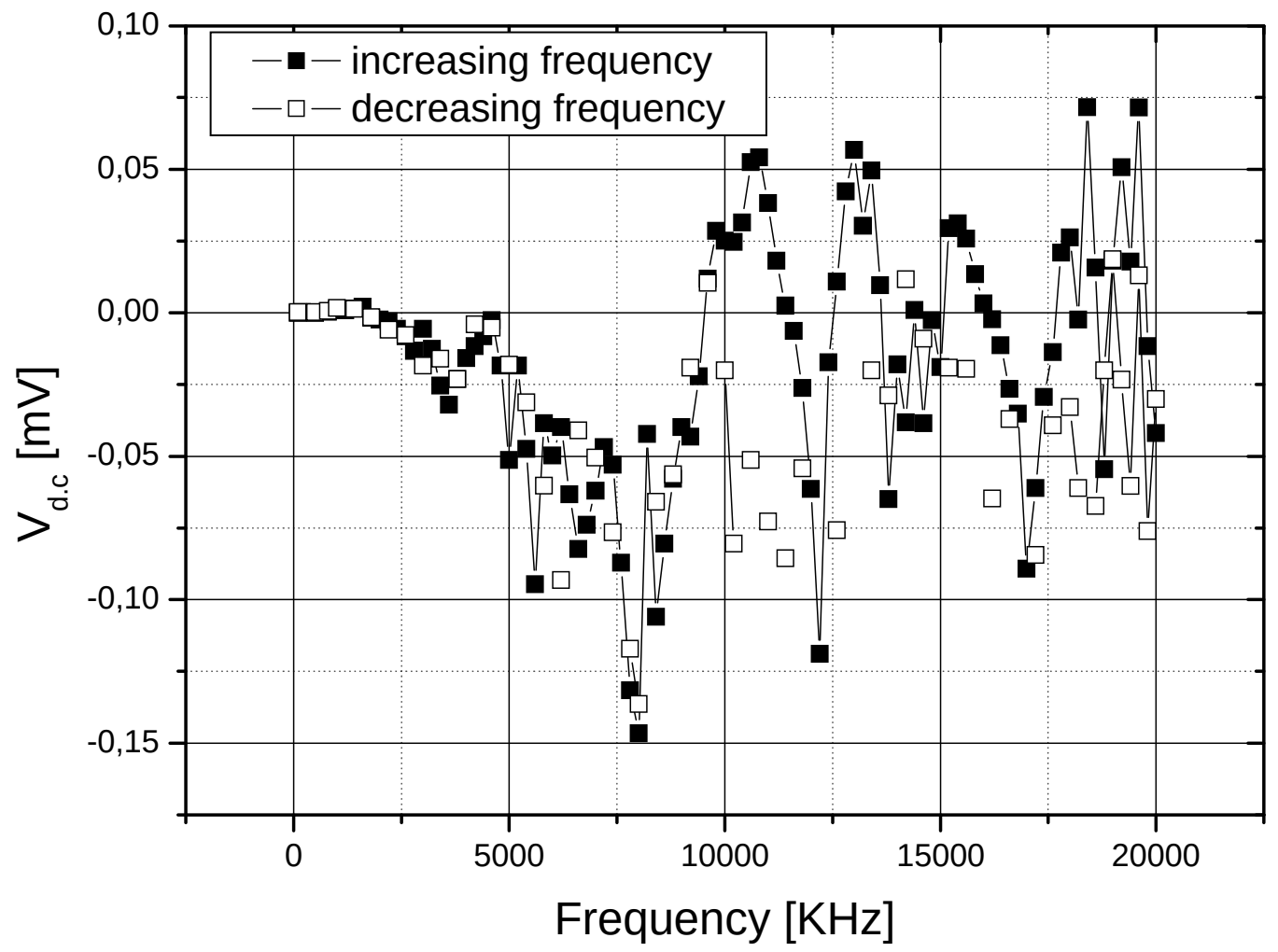

**Figure 3**

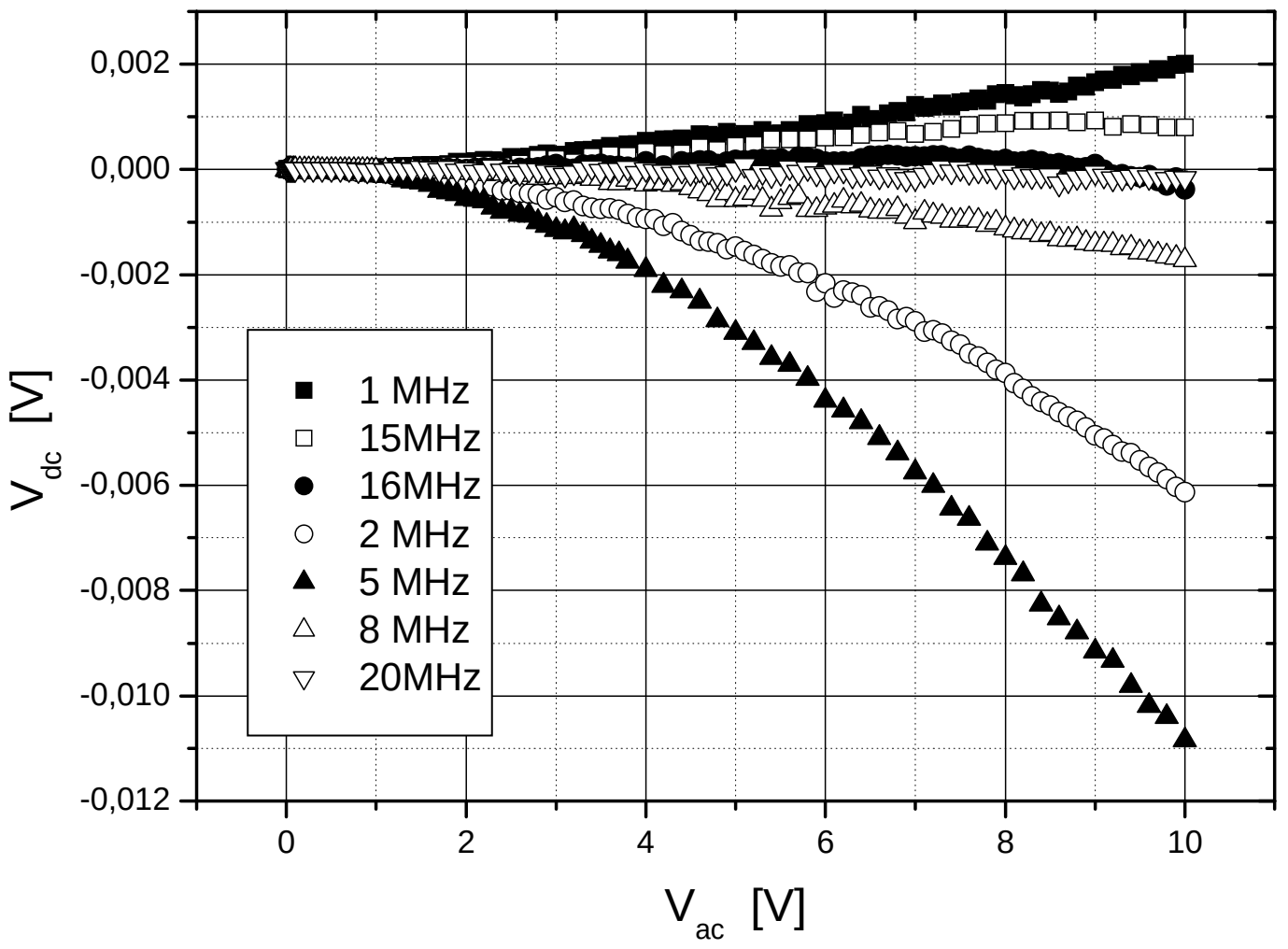

**Inset of Figure 3**

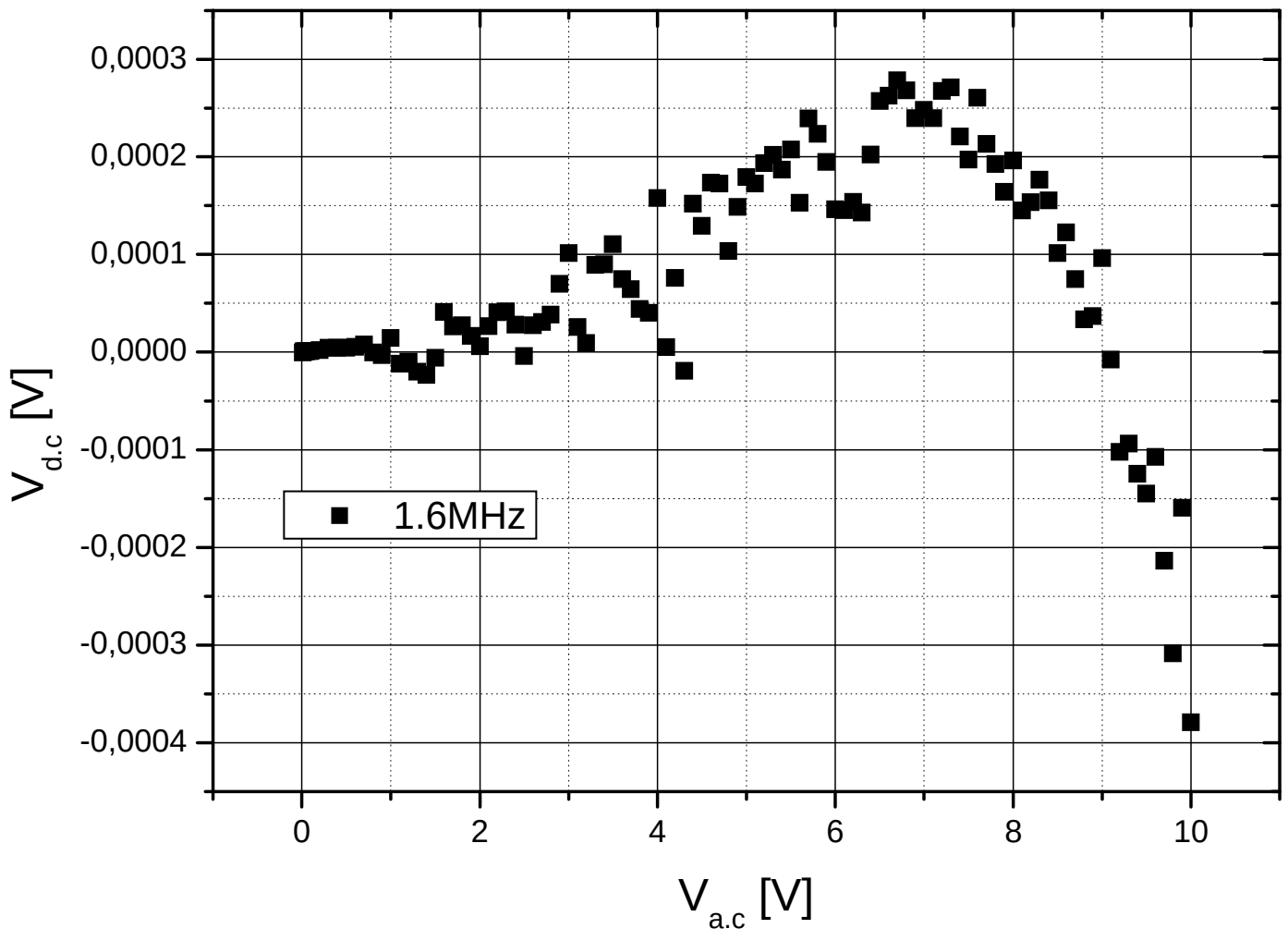

**Figure 4**

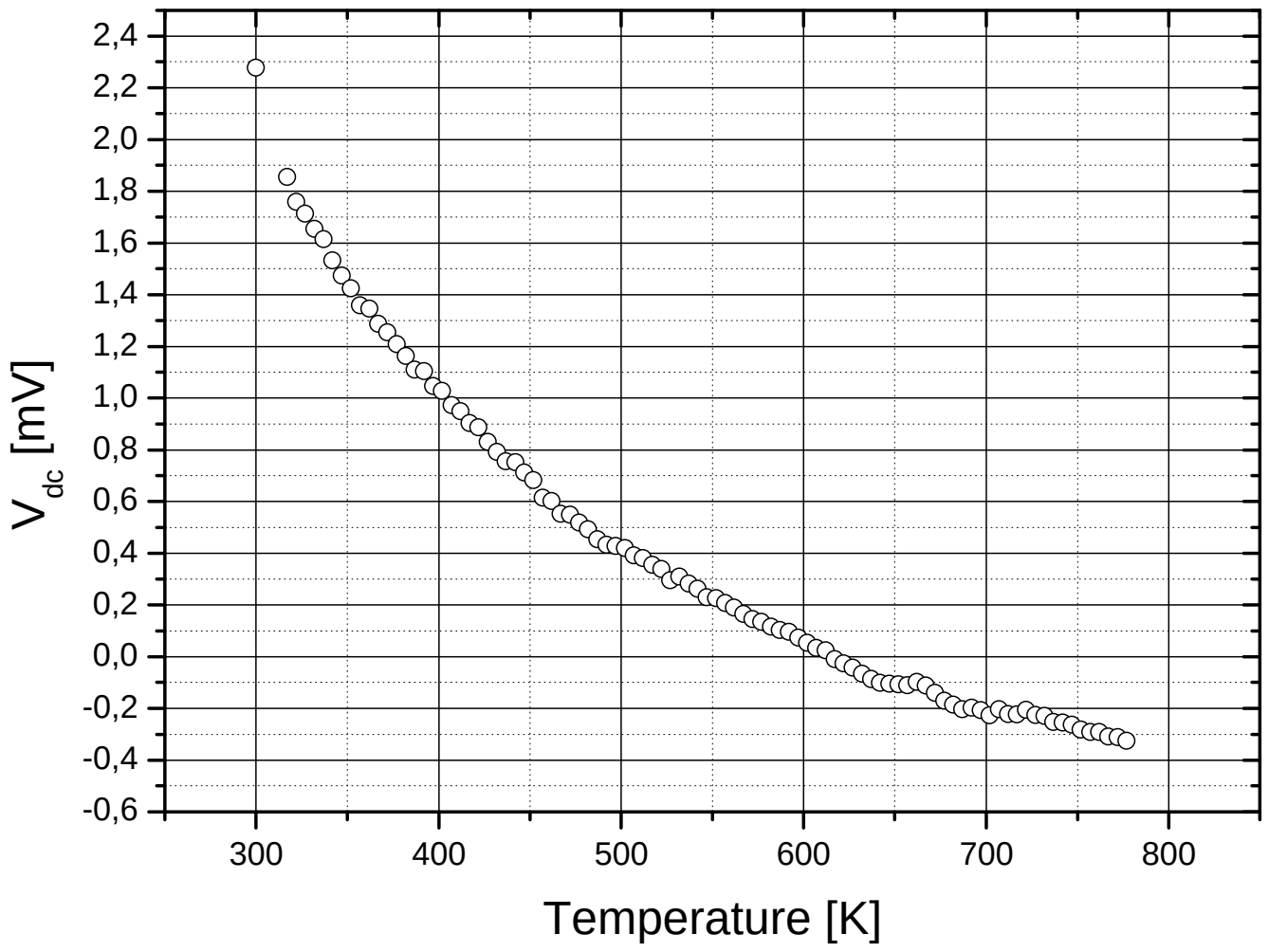

**Figure 5**

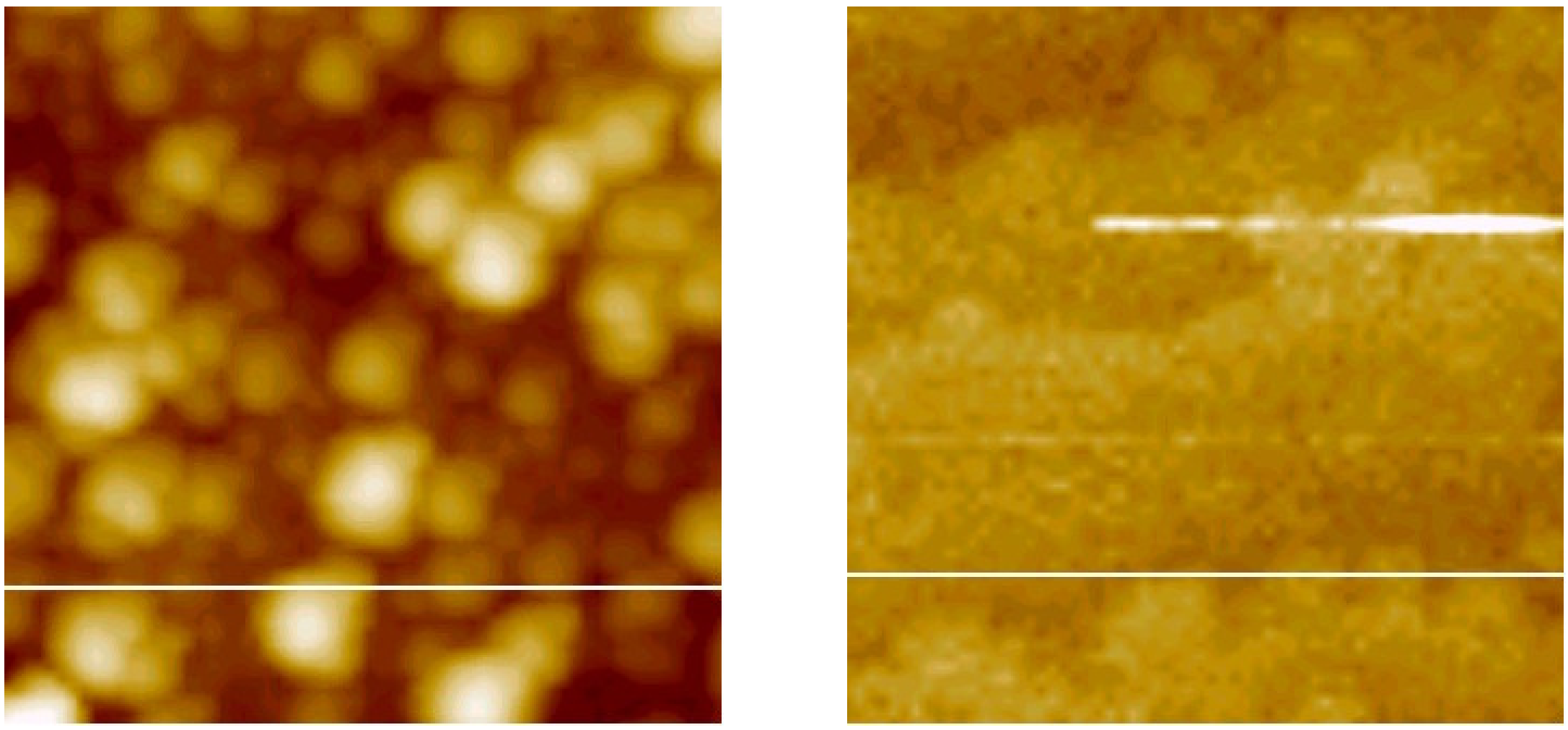

**Figure 6**

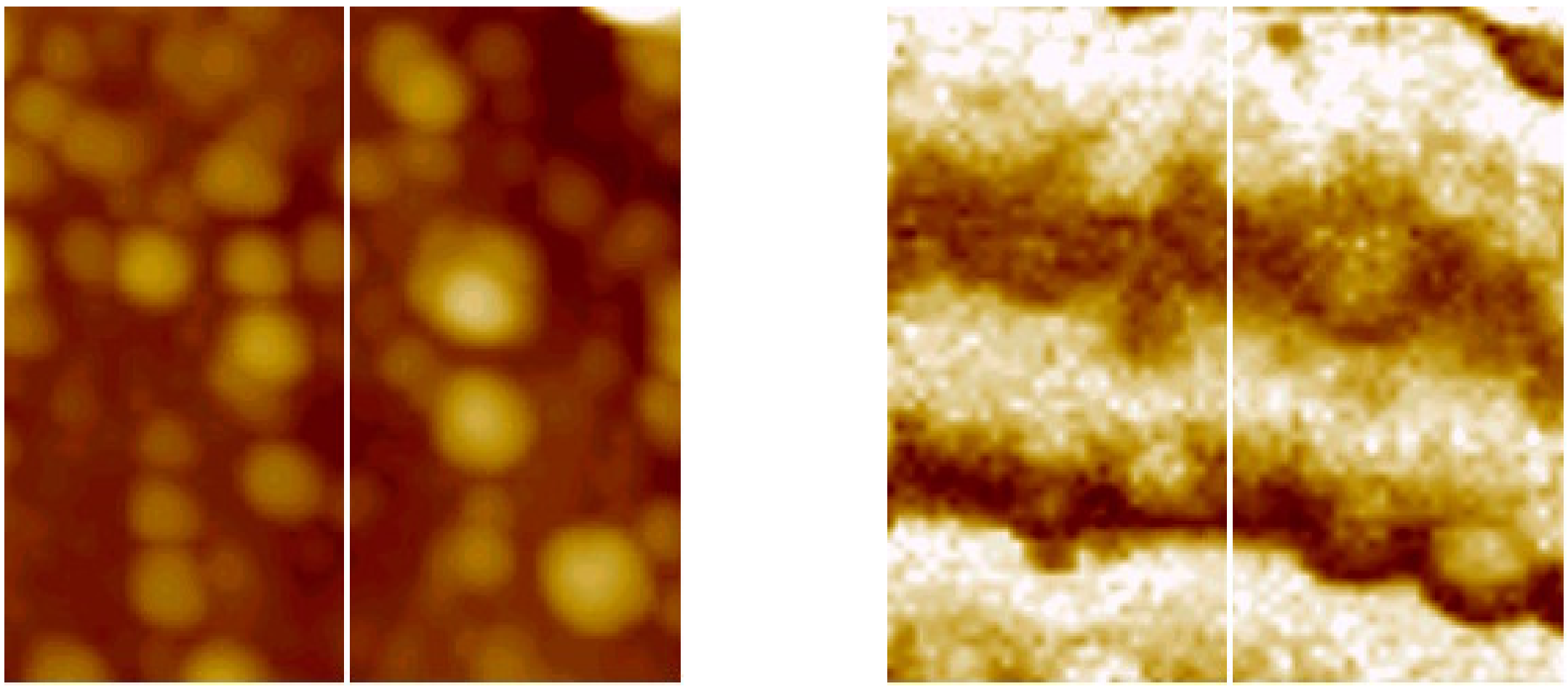

**Figure 7**

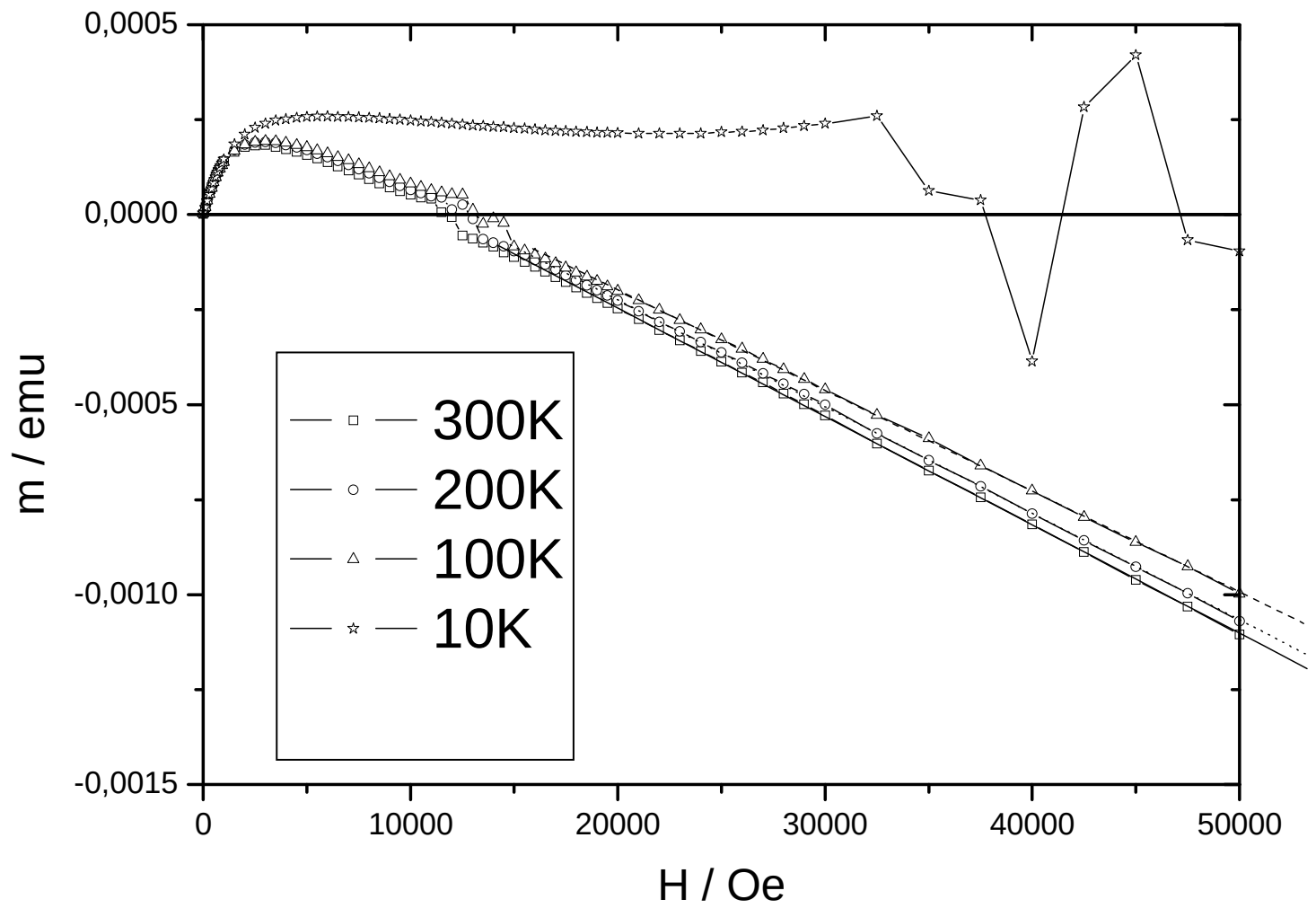

## Inset of Figure 7

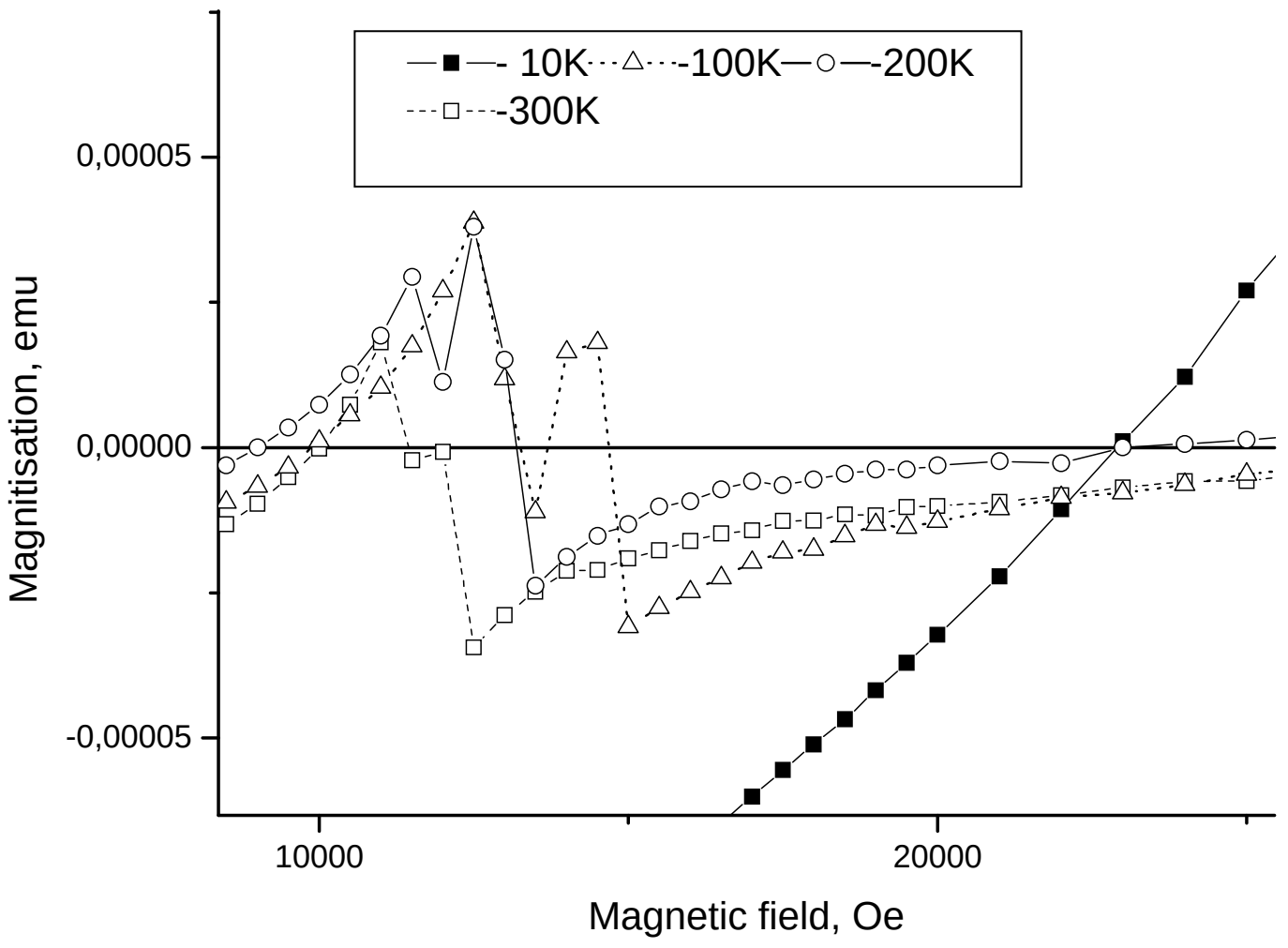